\documentstyle{article}

\begin{document}
\large
\begin{flushright}
ITP.SB-93-22\\
\end{flushright}

\Large
\begin{center}
{\bf  High-Order Adiabatic Approximation for Non-Hermitian Quantum System
and Complexization of Berry's Phase}
\footnote{\Large accepted for publication in {\it Physica Scripta}}
\vspace{1cm}

Chang-Pu Sun\\
Physics Dpartment, Northeast Normal University,
Changchun 130024, P.R.China\\
and\\
Institute for Theoretical Physics, State University of New York, Stony Brook,
NY 11794-3840, USA\\

\vspace{1cm}

{\bf Abstract}\\
\end{center}

\large
In this paper the evolution of a quantum system drived by a non-Hermitian
Hamiltonian depending on slowly-changing parameters is studied by building
an universal high-order adiabatic approximation(HOAA)
 method with Berry's phase ,
which is valid for either the Hermitian or the non-Hermitian cases
. This method can be regarded as a non-trivial generalization of the HOAA
method
for closed quantum system presented by this author before. In a general
situation, the probabilities of adiabatic decay and non-adiabatic transitions
are explicitly obtained for the evolution of the non-Hermitian quantum system.
It is also shown that the non-Hermitian analog of the Berry's phase factor
for the non-Hermitian case just enjoys the holonomy structure of the
dual linear bundle over the parameter manifold. The non-Hermitian evolution
of the generalized forced harmonic oscillator
is discussed as an illustrative examples.
\newpage
{\bf 1.Introduction}
\vspace{0.4cm}

Since Berry's phase factor(BPF) was discovered in evolution of the quantum
system with adiabatically-changing parameters [1], a few methods studying
non-adiabatic evolution of a quantum system driven by a Hermitian Hamiltonian
depending on slowly (but not adiabatically)-changing
parameters have been presented
in connection with BPF's [2-5]. In these methods, the high-order adiabatic
approximation (HOAA) method was proposed by this author for the first time [4]
and has been used and developed for many case [6-12]. However, all of these
studies has not been concerned with a kind of important quantum system, the
quantum open system that possesses a non-Hermitian (nH) Hamiltonian. This
paper  will be devoted to the generalization of the HOAA method for such nH
quantum system.

In fact, though the Hamiltonian for a closed quantum system, which is usually
considered as a basic object in quantum theory, must be a Hermitian operator,
many theories,  such as the Fock-Krylov theorem [13], show the probability to
apply the nH Hamiltonian for those quantum phenomena with dissipation, decay
and relaxation [14,15]. Recently, many practical problems including the
multiphoton ionization, the supermode free-electron laser and the transverse
mode propagation in optical resonator [16-19] have been concerned with the
use of nH Schrodinger equation and nH Hamiltonian correspondingly. Since the
occurrence of a BPF or its analog may be established when something in the
considered system is varied, it is natural to generalize the concept of BPF
for nH quantum system. More recently, some authors made this generalization
and applied it to concrete physical problems [20-21], but their studies were
only focused on the adiabatic case that the parameters change so slowly that
transition between any two (quasi-) energy levels do not happen. In this paper
we especially emphasized the non-adiabatic evolution of the nH quantum system
and the geometry of the nH analog of BPF.

The paper is arranged as follows. In section 2, using a similarity
transformation of the nH  Hamiltonian, we build an universal formalism of the
HOAA method for the nH quantum system. It is also available to the
Hermitian quantum system. In section 3, we explicitly analyse the conditions
under which the lowest order approximation,namely the the adiabatic
approximation,can work well. We also compare our results with that obtained
with the biorthonormal state method [20-21] in adiabatic case. In section 4,
we apply the general result to a simple toy model - the nH forced oscillator to
 show the usefulness of our analysis. In section 5, we show that the nH analog
of BPF appearing in an adiabatic evolution of nH quantum system are non-unitary
holonomy group element for the dual bundles  over the parameter maniford. In
the
further studies, we will provide more applications of this  generalized HOAA
method to some physical problems.
\vspace{0.4cm}

{\bf 2.Generalized HOAA Method for nH Quantum System}
\vspace{0.4cm}

Let us begin by setting some notations. The Hamiltonian of the quantum open
system  we consider as fillows is a nH operator
$${\bf H}={\bf H}(t)={\bf H}[{\bf R}]={\bf H}[{\bf R}(t)]$$
that depends on a set of slowly-changing parameters
$${\bf R}={\bf R}(t):(R_1(t),R_2(t),...,R_L(t)).$$
We now assume that {\bf H}(t) is diagonalizable at each instant t, i.e., there
exists a similarity transformation
$$U(t)=U[{\bf R}]=U[{\bf R}(t)]$$
such that
$$U(t){\bf H(t)}U(t)^{-1}=H_d(t)=\left[\begin{array}{lccr}
\epsilon_1(t) & 0      & ..... & 0 \\
0      &\epsilon _2(t) & ..... & 0 \\
...    &...     & ..... & ... \\
0      &     0  & ..... &\epsilon _N(t)
\end{array} \right ] = {\bf H}_d(t) \eqno{(2.1)}$$
where  the ``quasi-energy levels''
$$\epsilon_k(t) = \epsilon_k[{\bf R}] , ~~k=1,2...,N $$
may be complex and U(t) is not unitary correspondingly.

Let
$$|\Psi(t)>=U(t)|\Phi(t)>$$
be a solution of the nH Schrodinger equation
$$i\hbar \frac{\partial}{\partial t}|\Psi(t)>={\bf H}(t)|\Psi(t)>.
\eqno{2.2)}$$
The equivalent wavefunction $|\Phi(t)>$ must satisfy an equivalent Schrodinger
equation (ESE)
$$i\hbar \frac{\partial}{\partial t}|\Phi(t)>={\bf H}_e(t)|\Phi(t)>.
\eqno{2.3)}$$
with the equivelent Hamiltonian
$${\bf H}_e(t)={\bf H}_d(t)-i\hbar U(t)^{-1}\frac{\partial U(t) }{\partial t }
,\eqno{(2.4)}$$
\vspace{0.4cm}

Separate ${\bf H}_e(t)$ into two parts, the diagonal part
$${\bf H}_0(t)={\bf H}_d(t)+ diagonal~~part~~of ~~[ -i\hbar U(t)^{-1}
 \frac{\partial}{\partial t}U(t) ] $$
and the off-diagonal part
$${\bf V}(t)= ~~off~diagonal~~part~~of~~[ -i\hbar U(t)^{-1}
 \frac{\partial}{\partial t}U(t) ] . $$
Then,
$${\bf H}_e(t)={\bf H}_0(t)+{\bf V(t)}.$$

Late on we will show that the diagonal part ${\bf H}_0(t)$ governs an
adiabatic evolution of the nH quantum system while the off-diagonal
part ${\bf V}(t)$ governs the non-adiabatic transitions among the
quasi-energy levels. In fact, since ${\bf V}(t)$ completely vanishes when
${\bf H}$ is independent of time t and
${\bf H}={\bf H}(t)={\bf H}[{\bf R}]$ is a smooth function of ${\bf R}(t)$
, we can regard  ${\bf V}(t)$ as a pertubation when ${\bf H}$ or ${\bf R}$
depends on time ``slightly'', i.e.,  ${\bf R}(t)$ changes as t slowly enough.

Based on the above  decomposation of ${\bf H}_e(t)$, we can use the standard
time-dependent pertubation theory to solve the ESE (2.3), obtaining a
approximate series solution
$$|\Phi(t)>=|\Phi^0(t)>+|\Phi^1(t)>+|\Phi^2(t)>+...+|\Phi^l(t)>+...,\eqno{(2-5)}$$
Here, the  $l'$th order solution $|\Phi^l(t)>$ is determined by
$$i\hbar \frac{\partial}{\partial t}|\Phi^0(t)>={\bf H}^0(t)|\Phi_0(t)>;$$
$$i\hbar \frac{\partial}{\partial t}|\Phi^l(t)>={\bf H_0}(t)|\Psi(t)>
+{\bf V}(t)|\Phi^{l-1}(t)>.\eqno{(2.6)}$$
$$l = 1,2,...,$$

Because the above equations of  $l'$th order solution $|\Phi^l(t)>$
only concern $(l-1)'$th order solutions $|\Phi^{l-1}(t)>$ , we  can
get each order approximate solution starting from the zero'th order one
$$|\Phi^0(t)>=exp[\frac{1}{i\hbar}\int^t_0{\bf H}_0(t')dt'  ]|\Phi(0)>$$
$$|\Phi(0)>=U(0)^{-1}|\Psi(0)>.\eqno{(2.7)}$$

In order to write out thse solutions in a quite explicit form, we use the
eigenstates
$$|1>=\left[\begin{array}{c}
 1\\
 0\\
 .\\
 .\\
 0\\
 0\\
\end{array} \right ],~~~
|2>=\left[\begin{array}{c}
 0\\
 1\\
 .\\
 .\\
 0\\
 0\\
\end{array} \right ]~~ , ... ,~~
|N-1>=\left[\begin{array}{c}
 0\\
 0\\
 .\\
 .\\
 1\\
 0\\
\end{array} \right ] ,~~~
|N>=\left[\begin{array}{c}
 0\\
 0\\
 .\\
 .\\
 0\\
 1\\
\end{array} \right ]$$
of ${\bf H}_0(t)$ with the corresponding eigenvalue
$$E_n(t)=\epsilon_n(t)-\hbar \gamma_n(t).$$
where the additional term to the quasi-energy
$$\gamma_n(t)=i\int^t_0 <n|U(t')^{-1}\frac{\partial U(t') }{\partial t'}
|n>dt'\eqno{(2.8)}$$
will be proved to be the nH analog of Berry's phase. Then, each order solution
$$|\Phi^l(t)>=\sum _{n=1}^N C_n^l(t)e^{i\gamma_n(t)}e^{\frac{1}{i\hbar}
\int^t_0\epsilon_n(t')dt'}|n>,\eqno{(2.9)}$$
follows from eq.(2.6) immediately. Here, the coefficients $ C_n^l(t)$
satisfy
$$ C_n^0(t)=<n|\Phi(0)>=<n|U(0)^{-1}|\Psi(0)>,$$
$$  C_n^l(t)=\sum _{m=1}^N \frac{1}{i\hbar}\int^t_0<n|{\bf V}(t')|m>
e^{i\int^{t'}_0\omega_{n,m}(s)ds}C_m^{l-1}(t')dt',\eqno{(2.10)}$$
where
$$\omega_{n,m}=\frac{\epsilon_n(t)-\epsilon_m(t)+\dot{\gamma}_m(t)
-\dot{\gamma}_n(t)}{\hbar}$$

Notice that the difficulty that the eigenstates of ${\bf H}$ are not
orthogonal to each other  due to non-Hermiticity of {\bf H} has been avoided
in the above discussion by a trick building the ESE (2.3) to find the
pertubation {\bf V}(t). This is the key to our studies in this paper.
\vspace {0.4cm}

{\bf 3.Adiabatic Approximation and Comparision with Biorthonormal State Method}
\vspace {0.4cm}

In this section our focus first is on the adiabatic conditions that the
zero'th order (adiabatic) approximation can well approach the true evolution
of the nH quantum system. Consider the integral

$${\bf I}=\int^t_0<n|{\bf V}(t')|m>e^{i\int^{t'}_0\omega_{n,m}(s)ds}dt',$$
$$ =\int^t_0<n|{\bf V}(t')|m>e^{-Im[\int^{t'}_0\omega_{n,m}(s)ds]}
e^{iRe[\int^{t'}_0\omega_{n,m}(s)ds]}dt'
$$
appearing in the first order approximation

$$  C_n ^1(t)=\sum _{m=1}^N \frac{1}{i\hbar}<n|\Phi(0)>
\int^{t'}_0<n|{\bf V}(t')|m>
e^{i\int^{t'}_0\omega_{n,m}(s)ds}dt'.$$
Here, we have separated the damping factor
$$e^{-Im[\int^t_0\omega_{n,m}(s)ds]} $$
and the oscillating factor
$$e^{iRe[\int^t_0\omega_{n,m}(s)ds]}.$$

If the latter oscillates so fast that the conditions
$$|\frac{<n|{\bf V}(t)|m>e^{-Im[\int^t_0\omega_{n,m}(s)ds]}}
{Re[\omega_{n,m}(t)]\hbar}|\ll 1\eqno{(3.1)}$$
hold for $m\neq n$, the integral {\bf I}  tends to zero. This statment
can be proved by integrating {\bf I} by part. Then, we get so-called adiabatic
conditions (3.1). When they hold for a nH quantum system, we can ignore the
higher-approximation solutions $|\Phi_l(t)>,(l=1,2,...)$ .

To compare our analysis with the biorthonormal state method in refs.[20,21]
in the adiabatic case, we define
$$|\phi_n(t)>=|\phi_n[{\bf R}]>=U(t)|n>,$$
$$|\chi_n(t)>=|\chi_n[{\bf R}]>=[U(t)^{-1}]^\dagger|n>,\eqno{(3.2)}$$
Obiviously, they are the eigenstates of ${\bf H}(t)$ and ${\bf H}(t)^\dagger$
respectively with the eigenvalues $\epsilon_n(t)$ and $\epsilon^*_n(t)$.
Using the completeness relation and the orthonormal relations of states
$|n> (n=1,2,...,N)$:
$$\sum^N_{n=1}|n><n|=I(unit~~operator),~~~~~~~<n|m>=\delta_{m,n},$$
we immmediately get the generalized completeness relations
$$ \sum^N_{n=1}|\chi_n(t)><\phi_n(t)|=\sum^N_{n=1}|\phi_n(t)><\chi_n(t)|
=I\eqno{(3.3)} $$
and the biorthonomarl relations
$$<\phi_m(t)\chi_n(t)>=<\chi_m(t)|\phi_n(t)>=\delta_{m,n}.\eqno{(3.4)} $$

In terms of the biorthonormal basis $\{|\phi_n(t)>,  |\chi_n(t)> \}$,
we reexpress the high-order approximations of the nH quantum evolution as
$$|\Psi(t)>=\sum _{l=0}^{\infty}|\Psi^l(t)>=\sum _{l=0}^{\infty}U(t)
|\Phi^l(t)>$$
$$=\sum _{l=0}^{\infty}\sum _{n=1}^N C_n^l(t)e^{i\gamma_n(t)}e^{\frac{1}
{i\hbar}
\int^{t'}_0\epsilon_n(t')dt'}|\phi_n(t)>,\eqno{(3.5)}$$
with the coefficient equations:
$$ C_n^0(t)=<\chi_n(0)|\Psi(0)>,$$
$$  C_n^l(t)= - \sum _{m=1}^N \frac{1}{i\hbar}\int^t_0<\chi_n(t')|\phi_m(t')>
e^{i\int^{t'}_0\omega_{n,m}(s)ds}C_m^{l-1}(t')dt',\eqno{(3.5')}$$
for $l=1,2,3, ...$. Here, the nH analog of Berry's phase is rewritten as
$$\gamma_n(t)=i\int^t_0<\chi_n(t')| \frac{d}{dt}  \phi_n(t')>
dt'\eqno{(3.6)}$$
Correspondingly, the adiabatic conditions (3.1) are rewritten as

$$|\frac{<\chi_n(t)| \frac{d}{dt'}  \phi_n(t)>
e^{-Im[\int^t_0\omega_{n,m}(s)ds]}}
{Re[\omega_{n,m}(t)]}|\ll 1\eqno{(3.7)}$$

Now, let us consider the adiabatic evolution of an instantaneous eigenstate of
{\bf H}(t) under the adiabatic conditions. if the system is initially in state
$$|\Phi(0)>=U(t=0)|n>=|\phi_n(0)>,$$
it will evolve into
$$|\Phi(t)>=e^{i\gamma_n(t)}e^{\frac{1}{i\hbar}\int^t_0\epsilon_n(t')dt'}
|\phi_n(t)>$$
$$\equiv e^{i\Omega_n(t)}|\phi_n(t)>\eqno{(3.8)} $$
where
$$\Omega_n(t)=\gamma_n(t)-\frac{1}{\hbar}\int^t_0\epsilon_n(t')dt'.$$
The above eq. (3.8) manifests that the adiabatically- evolving state
is always an eigenstate of  the instantaneous Hamiltonian {\bf H}(t) if
the initial state is such state at t=0. This is the quantum adiabatic theorem
for nH quamtum systems, which is quite similar to that for the Hermitian case.
It shows the quantum number labelling the quasi-energy level to be an
adiabatic invariant. Since $\gamma_n(t)$ and $\epsilon_n(t)$ are usually not
real due to the non-Hermiticity of {\bf H}(t), the damping factor
$exp[-Im \Omega_n(t)]$ causes the adiabatic decay of the state. Especially,
when $\epsilon_n(t)$ is real, the decay only results from the nH-analog
of BPF and is a purely geometrical effect. In next section an example will
be used to illustrate this situation.

When the adiabatic conditions (3.7) are broken, there may appear the
transitions from an instantaneous eigenstate $|\phi_n(0)>$ to others
$|\phi_n(t)>$ for $m\neq n$. From eqs.(3.5) or (2.5,2.9,2.10), we obtaine the
transition probabilities
$$P(n\rightarrow m)=\frac{1}{\hbar^2}| \int^t_0<m|{\bf V}(t')|n>
e^{i\int^{t'}_0\omega_{m,n}(s)ds}dt'|^2e^{-2Im[\Omega(t)]}\eqno{(3.9)}$$
or
$$P(n\rightarrow m)=\frac{1}{\hbar^2}| \int^t_0<\chi_m(t')| \frac{d}{dt'}
\phi_n(t')>
e^{i\int^{t'}_0\omega_{m,n}(s)ds}dt'|^2e^{-2Im[\Omega(t)]}.\eqno{(3.9')}$$
Therefore, we reach the statement mentioned in section.2 that the diagonal
and off-diagnoal parts respectively describe the adiabatic evolution
and the non-adiabatic transitions for the quantum system. Notice that Berry
also
discussed the geometrical phase effect in non-adiabatic transition in
connection
with the non-Hermiticity due to the complexization of time [22].
\vspace{0.4cm}

{\bf 4.Example:Generalized Force Hamornic Oscillator}
\vspace{0.4cm}

Now, let us apply the above general analysis to a simple model - the
generalized
forced harmonic oscillator. The time-dependent Hamiltonian for this model is
$${\bf H}(t)={\bf H}[\alpha,\beta]=\hbar \omega (a^\dagger a +
\beta a^\dagger + \alpha a)\eqno{(4.1)}$$
where $\alpha=\alpha(t)$ and $ \beta=\beta(t)$ are slowly-changing complex
parameters, the constant $ \omega$ is real , $a$ and $a^\dagger$
are respectively the creation and annihilation operators for a  boson
state, which satisfy
$$[a , a^\dagger]=1.$$
For the usual forced harmonic oscillator (FHO), there is constraint
$\alpha=\beta^*$ and the
corresponding Hamiltonian is Hermitian. However, we make here a generalization
to
remove this constraint so that the Hamiltonian (4.10) is non-Hermitian.

Using the translated boson operators
$$A(\alpha)^\dagger = a^\dagger+\alpha ,~~A(\beta)=a + \beta \eqno{(4.2)}$$
obeying the same boson commutation relation as the above
$$[A(\beta) , A(\alpha)^\dagger ]=1,$$
we rewrite down the Hamiltonian (4.1) as
$${\bf H}(t)={\bf H}[\alpha,\beta]=\hbar \omega [ A(\alpha)^\dagger A(\beta)
-\alpha\beta].$$
Based on the above expression for {\bf H}(t), we immediately obtain
the instantaneous eigenstates of ${\bf H}(t)={\bf H}[\alpha,\beta]$
$$|\phi_n(t)>=|\phi_n[\alpha,\beta]>=|n[\alpha,\beta]>$$
$$=\frac{1}{\sqrt{n!}}[ A(\alpha)^\dagger]^n |0(\beta)>\eqno{(4.3)}$$
with the eigenvalues
$$\epsilon_n(t)=\epsilon_n[\alpha,\beta]=(n-\alpha\beta)\hbar \omega.$$
Here, the translated vacuum state $ |0(\beta)>$ obeys
$$A(\beta)|0(\beta)> = 0 , ~or~~ a|0(\beta)>=-\beta |0(\beta)>,$$
that is to say, $ |0(\beta)>$ is a coherent state
$$|0(\beta)>=C_{\alpha,\beta}e^{-\beta a^\dagger}|0>,\eqno{(4.4)}$$
where $|0>$  is the usual vacuum state satisfying  $a |0> = 0$ and
$C_{\alpha,\beta}$ is to be determined by the biorthonormal conditions
. According to the above analysis, we can finally write the explicit form
of $|\phi_n(t)>$
$$|\phi_n(t)>=C_{\alpha,\beta}\sum^n_{k=0}
\frac{n! \beta^{n-k}}{(n-k)!\sqrt{k!}}e^{-\alpha a^\dagger}|k>,\eqno{(4.3')}$$
where
$$|k>=\frac{1}{\sqrt{k!}}[a^\dagger]^k |0>$$
is the k boson state.

Noticing
$${\bf H}[\alpha,\beta]^\dagger={\bf H}[\beta^* , \alpha^*],$$
we easily obtain the dual vector $|\chi_n(t)>$ to $|\phi_n(t)>$:
$$|\chi_n(t)> = |\chi_n[\alpha,\beta]>=|n[\beta^* , \alpha^*]>$$
as the  eigenvectors of ${\bf H}[\alpha,\beta]^\dagger$ with the corresponding
eigenvalue $\epsilon_n[\beta^* , \alpha^*]$. Then, the biorthonormal conditions
$$<\chi_n(t)|\phi_m(t)>=\delta_{m,n}$$
define the biorthonormalization coeeficient
$$C_{\alpha,\beta}=e^{-\frac{1}{2}\alpha\beta}.$$

Taking into acount the translation transformation
$$e^{\xi a} a^\dagger e^{-\xi a}=a^\dagger+\xi, $$
$$e^{\xi a^\dagger} a e^{-\xi a^\dagger}=a - \xi,\eqno{(4.5)}$$
$$ \xi \in complex~~field~{\it C}$$
we get the U(t)-operator expression of the biorthonormal basis
$\{|\phi_m(t)> , |\chi_n(t)> \}$
$$|\phi_n(t)>=|\phi_n[{\bf R}]>=U(t)|n>,$$
$$|\chi_n(t)>=|\chi_n[{\bf R}]>=[U(t)^{-1}]^\dagger|n>,\eqno{(4.6)}$$
where the U(t)-operator
$$U(t)=U[\alpha,\beta]=e^{-\frac{1}{2}\alpha\beta}e^{-\beta a^\dagger}
e^{-\alpha a}\eqno{(4.7)}$$
diagonalizes the Hamiltonian {\bf H}(t), i.e.,
$$U(t)^{-1} {\bf H}(t) U(t)=\hbar \omega a^\dagger a .$$

Now, let us discuss the evolution governed by the nH Hamiltonian (4.1).
To this end, we first calculate
$$<\chi_m(t)|\frac{d}{dt}\phi_n(t)>=\sqrt n \dot{\alpha}\delta_{m,n-1}
-  \sqrt{ n+1} \dot{\beta}\delta_{m,n+1}$$
$$+\frac{1}{2}[\alpha\dot{\beta}-\beta \dot{\alpha}]\delta_{m,n}\eqno{(4.8)}$$
Then, we obtain the nH analog of BPF
$$e^{i\gamma_n(t)}\equiv e^{i\gamma(t)} =
exp\{\frac{i}{2}\int^t_0[\alpha\dot{\beta}-\beta \dot{\alpha}](t')dt'\}$$
$$=exp\{\frac{i}{2}\int^{{\bf R}(t)}_{C;{\bf R}(0)}
[\alpha d\beta-\beta d\alpha]\}\eqno{(4.9)}$$
where C is a curve \{{\bf R}(t) \} in a four dimensional parameter
manifold
$${\bf M :} \{{\bf R}=(\alpha , \beta) =(\alpha_1, \alpha_2,
\beta_1, \beta_2
 ) | \alpha=\alpha_1 + i\alpha_2, \beta=\beta_1 +i\beta_2 \}
$$
where $\alpha_1, \alpha_2,
\beta_1~~ and ~~\beta_2$ are real. Because the factor $e^{i\gamma_n(t)}$ is
independent of the quantum number n, the adiabatic evolution
$$|\Phi(t)>=e^{i\gamma(t)}\sum ^N_{n=1}
<\chi_n(0)|\Psi(0)>e^{-in\omega t}|\phi_n(t)>$$
$$\equiv e^{-\Gamma (t)} e^{i\Theta(t)}
\sum ^N_{n=1} <\chi_n(0)|\Psi(0)>e^{-in\omega t}|\phi_n(t)\eqno{(4.10)} $$
is accompanied with a global geometrical factor with the oscillating part
$ e^{i\Theta(t)}$ :
$$\Theta(t) = Re[\gamma_n(t)]= \frac{i}{2}\int^t_0
[\beta_1 \dot{\alpha}_2+\beta_2 \dot{\alpha}_1 - \alpha_2\dot{\beta}_1
-\alpha_1\dot{\beta}_2](t')dt'$$
and the damping part $ e^{-\Gamma (t)} $:
$$\Gamma (t) = Im[\gamma_n(t)]= \frac{i}{2}\int^t_0
[\beta_2 \dot{\alpha}_2-\beta_1\dot{\alpha}_1 - \alpha_2\dot{\beta}_2
+\alpha_1\dot{\beta}_1](t')dt'$$
Notice that, when $\alpha(t) = \beta (t)^*$ , the BPF of coherent state for FHO
in ref.[23] is given once again as a special case of eq.(4.9).

For the case with large quantum number n or rapidly-changing parameters
$\alpha(t)$ and  $\beta (t)$, the adiabatic conditions
$$\frac{\sqrt n |\dot{\alpha}(t)|}{\omega} e^{-\Gamma (t)} \ll 1;
\frac{\sqrt {n+1} |\dot{\beta}(t)|}{\omega} e^{-\Gamma (t)} \ll 1
\eqno{(4.11)} $$
do not hold and we need to consider the non-adiabatic effects caused by the
first order approximation at least. If the system is initially in a
state $|\phi_k(0)>$, the initial conditions
$$C_n^0(0)=\delta_{n,k}~~ ,~~ C_n^l(0)=0 , for ~~l\geq 1$$
leads to the first order approximate solution
$$|\Psi^1(t)>=\sqrt {k}e^{-\Gamma(t)} e^{i[\Theta(t)-(k-1)\omega t]}
\int^t_0 \dot{\alpha}(t') e^{-i\omega t']}dt'~|\phi_{k-1}(t)>$$
$$+\sqrt {k+1}e^{-\Gamma(t)} e^{i[\Theta(t)-(k+1)\omega t]}
\int^t_0 \dot{\beta}(t') e^{i\omega t'}dt'|\phi_{k+1}(t)> \eqno{(4.12)}$$
The probabilities of the transition from  $|\phi_n(0)>$ to
$|\phi_n(t)>$ (n=k-1, k+1) are
$$P(k\rightarrow n)=k e^{-2\Gamma(t)}|\int^t_0 \dot{\alpha}(t') e^{-i\omega t'}
dt'\ |^2\delta_{n,k-1} + (k+1)e^{-2\Gamma(t)}|\int^t_0 \dot{\beta}(t')
e^{i\omega t']}dt|^2\delta_{n,k+1}.\eqno{(4.13)}$$

Obviously, the selection rule for such transition is
$${\bf \Delta n = +1~~0r~~-1},\eqno{(4.14)}$$
under the first order approximation.
\vspace{0.4cm}

{\bf 5. The nH-Analog of BPF and Holonomy in Dual Line Fiber Bundles}
\vspace{0.4cm}

For Hermitian quantum system, Simon recognized that the BPF is precisely
the holonomy in a Hermitian line bundle defined by the adiabatic evolution.
The adiabatic evolution can be interpreted as a parallel translation in
such bundle [24]. Now, a question naturally arises for the non-Hermitian
case: what is a geometrical interpretation of the nH analog of BPF? The
answer will be given in this section.

Let {\bf M} be the parameter manifold formed by the  parameters
${\bf R}=(R_1,R_2,...,R_L) $ . A line bundle defined by the nH Hamiltonian
${\bf H}[{\bf R}]$ is
$${\bf F}_n=\{({\bf R}, |\sigma_n[{\bf R}]>)~~ |~~ {\bf H}[{\bf R}]
|\sigma_n[{\bf R}]>=\epsilon_n[{\bf R}] |\sigma_n[{\bf R}]>,{\bf R}
\in {\bf M}\} .$$
Its fiber space is a one-dimensional linear space
$${\bf V_n} :\{|\sigma_n[{\bf R}]>=e^{i\theta[{\bf R}]|}\phi_n[{\bf R}]> |
{}~~\theta[{\bf R}] ~is ~a~ real ~functions ~ depending ~ on ~{\bf R}\}$$
Over the same base manifold {\bf M}, the dual bundel ${\bf F}_n^*$
is defined by

$${\bf F}^*_n=\{({\bf R}, |\sigma^*_n[{\bf R}]>)~~ |~~ {\bf H}[{\bf R}]
^\dagger |\sigma_n^*[{\bf R}]>=\epsilon^*_n[{\bf R}] |\sigma_n^*
[{\bf R}]>,{\bf R}
\in {\bf M}\} .$$
where the fiber space
$${\bf V_n^*} :\{|\sigma_n^*[{\bf R}]>=e^{i\theta'[{\bf R}]}|\chi_n[{\bf R}]> |
{}~~\theta'[{\bf R}]~ is~ a ~ real ~functions ~ depending ~ on ~{\bf R}\}$$
is the dual space of ${\bf V_n}$  and $|\chi_n[{\bf R}]>$ is the dual
elemet of $|\phi_n[{\bf R}]>$ , i.e.,
$$<\chi_n[{\bf R}]|\phi_n[{\bf R}]>=\delta_{m,n} .$$

Since the quantum number n labelling an eigenstate  $|\phi_n[{\bf R}]>$
of the instantaneous Hamiltonian  ${\bf H}[{\bf R}]$ is an adiabatic
invariant, we can assume that
$$|\sigma_n(t)>=|\sigma_n[{\bf R}]>=C_n[{\bf R}]|\phi_n[{\bf R}]>$$
is an evolution state in adiabatic case. Now, let us show that
the holonomy group elements of ${\bf F}_n$ ( ${\bf F}^*_n$) is the nH
analog of BPF for the adiabatic evolution while
$\{|\sigma_n(t)>|t\in [0 ,T]\}$ ( $\{|\sigma*_n(t)>|t\in [0 ,T]\}$ ),
as a curve in ${\bf F}_n$ (${\bf F}^*_n$) , is a horizontal lift of the
curve
$C: \{{\bf R}(t) |t\in [0 ,T]\}$
 on base manifold {\bf M}. To this end we consider a decomposition of a
tangent vector
$$\frac{d}{dt}|\sigma_n(t)> =\frac{d_v}{dt}|\sigma_n(t)>
+ \frac{d_h}{dt}|\sigma_n(t)>
\eqno{(5.1)}$$
where the vertical part along the fiber is
$$\frac{d_v}{dt}|\sigma_n(t)>=<\chi_n[{\bf R}]|\frac{d}{dt}|\sigma_n[{\bf R}]>
|\phi_n[{\bf R}]>$$
$$=(\frac{d}{dt}C_n[{\bf R}]+<\chi_n[{\bf R}|\frac{d}{dt}|\phi_n[{\bf R}]>
C_n[{\bf R}])|\phi_n[{\bf R}]>\eqno{(5.2)}$$
and the horizontal part orthogonal to the fiber is
$$\frac{d_h}{dt}|\sigma_n(t)>=\sum_{m\neq n}<\chi_m[{\bf R}]
|\frac{d}{dt}|\sigma_n[{\bf R}]>|\phi_m[{\bf R}]> .\eqno{(5.3)}$$
A parallel-translation implies
$$<\chi_n[{\bf R}|\frac{d_v}{dt}|\sigma_n(t)> = 0 ,\eqno{(5.4)}$$
which results in a one-form equation
$$d C_n[{\bf R}]+<\chi_n[{\bf R}|d \phi_n[{\bf R}]>
C_n[{\bf R}]= 0.\eqno{(5.5)}$$
Its solution
$$C_n(t)=C_n[{\bf R(t)}] = e^{i\gamma_n(t)}$$
just gives the nH analog of BPF again.

For a cyclic evolution that ${\bf R}(0)={\bf R}(t)$ and C is a closed curve,
the complex phase $\gamma_n(t)$ in the nH analog of BPF can be expressed as
an closed path integration
$$\gamma_n(T) = \gamma_n[C]=\oint_{C}{\bf A_n[R]}\eqno{(5.6)}$$
of a complex potential one-form
$${\bf A_n[R]}= i<\chi_n[{\bf R}]|d \phi_n[{\bf R}]>$$

Similarly, on dual bundel ${\bf F}^*_n$, the parallell-translation
condition
$$<\phi_n[{\bf R}]|\frac{d_v}{dt}|\sigma*_n(t)> = 0 $$
results in the dual nH analog of BPF with the geometric phase
$$\bar{\gamma}_n(t) = i \int^t_0<\phi_n[{\bf R}(t')]|\frac{d}{dt'}
 \chi_n[{\bf R}(t')]>
dt'.\eqno{(5.7)}$$

Obviously, the nH analog $exp[\gamma_n(t)]$ and its dual
$exp[\bar{\gamma}_n(t)]$ occuring in the  adiabatic evolution
are holonomy group elements on the line bundel  ${\bf F}_n$
and its dual ${\bf F}^*_n$ respectively. Except for the effect of dynamical
factors
$$f(t) = e^{\frac{1}{i \hbar} \int^t_0 \epsilon[{\bf R}(t')]dt'} ,
\bar{f}(t) = e^{\frac{1}{i \hbar} \int^t_0 \epsilon[{\bf R}(t')]^*dt'}$$
the adiabatic effect in quantum evolution are equivalent to the
parallel-translation on the  line bundels  ${\bf F}_n$ and  its dual
${\bf F}_n^*$ .
This is a circumstance
 similar to that for the Hermitian quantum system.
\vspace{0.8cm}

\large
{\bf Acknowledgements}
\vspace{0.4cm}

This work is supported by Cha Chi- Ming fellowship through the CEEC in State
University of New York at Stony Brook. The author is also supported in
part by the NFS of China and The  Fok Ying-Tung Education Foundation through
Northeast Normal University.

\newpage
{\bf References}
\vspace{0.4cm}

\begin{enumerate}
\item M.V.Berry, Proc.R.Soc.Lond.,A392 (1984),45; for a complete review, see
 {\it Geometric Phases in Physics} ,ed,by A.Shapere and F.Wilczek, World
Scientific,Sigapore, 1989.
\item M.V.Berry, Proc.R.Soc.Lond.,A414 (1987),31.
\item N.Nakagawa, Ann.Phys., 179(1987),145.
\item C.P.Sun, J.Phys.A,21(1988),1585; High Energy Phys.Nucl.Phys.,12(1988),
351; Phys.Rev.D, 41(1990),1318.
\item C.M.Bender and N.Papanicolaou,J.Phys.France, 49(1988),561; 1493.
\item C.P.Sun,Phys.Rev.D, 38(1988),2908.
\item C.P.Sun and M.L.Ge,Phys.Rev.D, 41(1990),1349; Commun.Theor.Phys.,
13(1990),63; C.P.Sun and L.Z.Zhang,Commun.Theor.Phys.12(1989) 479.
\item C.P.Sun, Chinese Phys.Lett.,6(19989),481.
\item Z.Y.Wu,Phys.Rev.A ,40(1989) 2184; 6852.
\item B.L.Markovski, in {\it Topological Phases in Quantum Theory,}  ed. by
 B.L.Markovsk and S.I.Vinitsky, World Scientific,Sigapore, 1989.
\item S.I.Vinitsky, V.L.Derbov, V.N.Dubovski , B.L.Markovskiand
Yu.P.Stepanovski,Sov.Phys.Usp. 33(1990),403 .
\item  C.P.Sun ,L.wang and M.L.Ge, Commun.Theor.Phys.,15(1991),427.
\item V.Fock and K.Krylov,Zh.Eksp.Fiz.17(1947),93.
\item E.Divies, {\it Quantum Theory of Open system}, Academic, New York, 1976.
\item P.Exner, {\it Open Quantum system and Feynman Integrals}, Reidel,
Dordrecht, 1985.
\item F.M.Faysal and J.V.Moloney, J.Phys.B, 14(19820, 3603.
\item A.Siegman, Optics Commu.,31(1979),369.
\item H.C.Baker,Phys.Rev.Lett.,50(1983),1579; Phys.Rev.A,30(1984),773.
\item G.Dattoli,A.Torre and R.Mignani, Phys.Rev.A,42(1990),1476; and refs.
therein.
\item C.Miniatura, C .Sire,J.Baudon and J.Bellissad, Europ.Phs.Lett.,13(19900,
199.
\item  G.Dattoli, R.Mignani and A.Torre, J.Phys.A,23(1990),5795..
\item M.V.Berry, Proc.R.Soc.Lond.,A,430,405;429(1990),61.
\item S.Chaturved,M.S.Sriram and V.Srinivasan, J.Phys.A,20(1987),L1071.
\item B.Simon,Phys.Rev.Lett.,51(1983),2167.

\end{enumerate}

\end{document}